\documentclass[12pt]{article}
\begin{document}
\renewcommand{\textfraction}{0}
\def\IR{{\rm I \kern-0.20em R}}
\def\psp{{\mathrm{H}}}
\def\psa{{\mathcal{H}}}
\def\prv{{\Theta}}
\def\bbbr{{\IR}}
\def\prl{{\theta}}
\def\col{{\mathcal{C}}}
\def\colset{C}
\def\mir{\Gamma}
\def\cnt{n}
\newtheorem{thm}{Theorem (\cite{DevGyoLug96})}
\newtheorem{prop}{Proposition}
\newtheorem{cor}{Corollary}
\newtheorem{rem}{Remark}
\newtheorem{defn}{Definition}

\title{Distributed Learning in Wireless Sensor Networks\footnote{This research was supported in part by the Army Research Office under Grant
DAAD19-00-1-0466, in part by Draper Laboratory under Grant IR\&D
6002, in part by the National Science Foundation under Grants CCR-0020524 and 
CCR-0312413, and in part by the Office of Naval Research under Grant No.
N00014-03-1-0102.  }} \author{\normalsize
Joel B. Predd\,\,\,\,\,\,\,\,\,\, Sanjeev R. Kulkarni\,\,\,\,\,\,\,\,\,\, H. Vincent Poor\\
\small Dept. of Electrical Engineering \\[-5pt] \small Princeton University\\[-5pt]Ê
\small Princeton, NJ 08544 \\[-5pt] \small
\{jpredd, kulkarni, poor\}@princeton.eduÊ}
\date{}
\maketitle
\thispagestyle{empty}
\begin{abstract}
The problem of distributed or decentralized detection and estimation in applications such as wireless sensor networks has often been considered in the framework of parametric models, in which strong assumptions are made about a statistical description of nature.  In certain applications, such assumptions are warranted and systems designed from these models show promise.  However, in other scenarios, prior knowledge is at best vague and translating such knowledge into a statistical model is undesirable.   Applications such as these pave the way for a nonparametric study of distributed detection and estimation.  In this paper, we review recent work of the authors in which some elementary models for distributed learning are considered.  These models are in the spirit of classical work in nonparametric statistics and are applicable to wireless sensor networks.

\end{abstract}Ê
\normalsize

\section{Introduction}
Wireless sensor networks have attracted considerable attention in
recent years \cite{AkySuSanCay02}.  Research in this area has focused on two separate
aspects of such networks:  networking issues, such as capacity,
delay, and routing strategies; and applications issues.  This paper
is concerned with the second of these aspects of wireless sensor
networks, and in particular with the problem of distributed inference.
Wireless sensor networks are {\it a fortiori} designed for the purpose
of making inferences about  the environments that they are sensing,
and  they are  typically characterized by limited communications
capabilities due to tight energy and bandwidth limitations. Thus,
distributed inference is a major issue in the study of such networks.
Distributed detection and estimation is a well-developed field
with a rich history.  Much of the work in this area has focused on
either parametric problems, in which strong statistical assumptions
are made \cite{Var96, BluKasPoo97,Vee01, LiWonHuSay02, KumZhaShe02, CosAay03, KotRamSay03, DonTonSad04}, or on traditional nonparametric formalisms, such as
constant-false-alarm-rate detection \cite{BarVar89}. In this paper, we consider
an alternative nonparametric approach to distributed inference
that is relevant to wireless sensor networks, namely, distributed
learning under communications constraints.

Although \cite{Sim03} advocated a learning theory approach to sensor networks, \cite{NguWaiJor04} is the first work to consider the classical model for decentralized detection in a nonparametric setting.  In the context of kernel methods commonly used in machine learning, the notion of a marginalized kernel is introduced in \cite{NguWaiJor04} to derive an efficient algorithm for designing a decentralized detection system based on a collection of training data.

A related area of research lies in the study of ensemble methods in machine learning; examples of these techniques include bagging, boosting, mixtures of experts, and others \cite{JacJorNowHin91, Bre96, FreSch97b, FreSchSinWar97a, KitHatDuiMat98}.  These techniques are similar to the problem of interest here in that they aggregate many individually trained classifiers. However, the focus of these works is on the statistical and algorithmic advantages of learning with an ensemble and not on the nature of learning under communication constraints.  Notably, \cite{KeaSeu95} considered an early model for learning with many individually trained hypotheses.

The models considered in this paper have been studied in detail in \cite{PreKulPoo04a,PreKulPoo04b}.  Here, we focus on the models and the main results and refer the reader to \cite{PreKulPoo04a,PreKulPoo04b} for a more thorough discussion and proofs of these results.  In particular, there is extensive work in nonparametric statistics that is closely related to the models considered in this paper; we touch on these connections throughout, but leave a more complete review and list of references to the full papers.

In Section 2, we review the classical model for learning in nonparametric statistics and explain where our work departs the classical model.  In Section 3, we discuss a model for distributed learning with distributed data under a family of simple communication models \cite{PreKulPoo04b}; we discuss our main results and connect them to related work in nonparametrics  In Section 4, we discuss a model for distributed learning with specialists \cite{PreKulPoo04c, PreKulPoo04a}. Similarly, we connect the main result with known results in nonparametrics.   Finally, we end with conclusions in Section 5.

\section{The Classical Learning Model \& Our Departure}Ê
Let us briefly review a standard model for learning in nonparametric statistics.  	For a thorough introduction to nonparametric statistics and classical centralized learning models, we refer the reader to \cite{DevGyoLug96,GyoKohKrzWal02}.

Let $X$ and $Y$ be $\mathcal{X}$-valued and $\mathcal{Y}$-valued random variables, respectively,  with a joint distribution denoted by $\mathbf{P}_{XY}$.   ${\mathcal{X}}$ is known as the feature, input, or observation space; ${\mathcal{Y}}$ is known as the label or output space.   Throughout, we will take ${\mathcal{X}}\subseteq\IR^d$ and consider two cases corresponding to binary classification (${\mathcal{Y}}=\{0,1\}$) and regression estimation (${\mathcal{Y}}=\IR$).  Of course, the decision-theoretic problem is to predict $Y$, given an observation $X$.

In parametric settings, one assumes prior knowledge of the distribution ${\mathbf{P}}_{XY}$.  Defining a loss function $l:{\mathcal{Y}}\times{\mathcal{Y}}\rightarrow\IR$, one designs a decision rule that achieves the minimal expected loss $L^{\star}=\inf_{g}{\mathbf{E}}\{l(g(X), Y)\}$.  In the binary classification setting, the criterion of interest is the probability of misclassification; we let $l(y,y^{\prime}) = 1_{\{y\neq y^{\prime}\}}$, the well-known zero-one loss.  The structure of the risk minimizing decision rule is well-understood \cite{DevGyoLug96}; let $g_B:{\mathcal{X}}\rightarrow\{0,1\}$ denote this Bayes decision rule.

In regression settings, we consider the squared error criterion;  we let $l(y, y^{\prime}) = |y-y^{\prime}|^2$.    It is well known that the regression function $\eta(x) = {\mathbf{E}}\{Y\,| X=x\}$ achieves the minimal expected loss.

In nonparametric settings, prior knowledge of the distribution ${\mathbf{P}}_{XY}$ is not available and thus, computing the Bayes rule or regression function is not possible.  Hope is not lost, for we are provided  $D_n=\{(X_i, Y_i)\}_{i=1}^{n}$, an independent and identically distributed (i.i.d.) collection of training data with $(X_i, Y_i)\sim \mathbf{P}_{XY}$ for all $i\in\{1,...,n\}$.  The learning problem is to use this data to infer decision rules with small loss.   That is, our decision rules are independent of ${\mathbf{P}}_{XY}$, but can depend on the labeled examples in $D_n$;  i.e., $g(X) = g(X, D_n)$.  

In this work, we focus on the information-theoretic property known as \emph{universal consistency} \cite{DevGyoLug96,GyoKohKrzWal02}.   Though a thorough discussion of this property is beyond the scope of this paper,  we will state the definition for the sake of continuity. 
\begin{defn}
Let $L_n={\mathbf{E}}\{l(g_n(X, D_n), Y)\,| D_n\}$. $\{g_n\}_{n=1}^{\infty}$ is said to be \emph{universally consistent} if $\mathbf{E}\{L_n\}\rightarrow L^{\star}$ for \emph{all} distributions ${\mathbf{P}}_{XY}$.
\end{defn}
Consistent with convention, we use $g_n(x) = g_n(x, D_n)$ to denote decision rules in the binary classification setting and we use $\hat{\eta}(x) = \hat{\eta}(x, D_n)$ to denote decision rules in the regression setting.

The existence of universally consistent classifiers and estimators was an open question until Stone's Theorem \cite{Sto77} demonstrated that a wide range of classifiers and estimators had this fundamental property; rules in this class are known as weighted-average rules and include histogram estimators, nearest-neighbor rules, and classical kernel rules.  Extensive work in nonparametrics has extended this result to consider the consistency of Stone-type rules under various sampling processes; see, for example, \cite{DevGyoLug96,GyoKohKrzWal02} and references therein.  These models focus on various dependency structures within the training data and assume that a single processor has access to the entire data stream.  

In distributed scenarios like sensor networks, different sensors have access to different data streams that differ in distribution and may depend on external parameters such as the state of a sensor network or location of a database.  Moreover, sensors are unable to share all of their data with each other or with a central fusion center, as they may have only a few bits with which to communicate a summary.  The nature of the work considered in this paper is to consider questions of universal consistency similar to those above but in this distributed environment.  For a given model of communication amongst sensors, each of whom has been allocated a small portion of a larger learning problem,  can enough information can be exchanged to allow for a universally consistent network? We consider several models that differ both in the way the learning problem is distributed amongst sensors and in the nature of the communication constraints.  These models more closely resemble a distributed environment and present  new questions to consider with regard to universal consistency.  Insofar as these models present a useful picture of distributed scenarios, this paper addresses the issue of whether or not the guarantees provided by Stone's Theorem in centralized environments hold in distributed settings.  Notably, the models under consideration will be similar in spirit to their classical counterparts;  indeed, similar techniques can be applied to prove results.

\section{Learning with Distributed Data}

\subsection{The Model }
In this section,  we present a model where the learning problem is divided amongst sensors by distributing examples from an i.i.d. training set amongst the sensors.  In the classical setting, the training data $D_n$ is provided to a single, centralized learning agent.  Instead, suppose that for each $i\in\{1,...,n\}$, the training datum $(X_i, Y_i)$ is received by a distinct member of a network of $n$ sensors.  When the fusion center observes a new observation $X\sim\mathbf{P}_X$,  it broadcasts the observation to the network in a request for information.  At this time, each sensor can respond with at most one bit.  That is, each sensor chooses whether or not to respond to the fusion center's request for information; if it chooses to respond, a sensor sends either a $1$ or a $0$ based on its local decision algorithm.  Upon observing the response of the network, the fusion center combines the information to create an estimate of $Y$.  As before, the key question is:  do there exist sensor decision rules and a fusion rule that result in a universally consistent network in the limit as the number of sensors increases without bound?

In Sections 3.2 and 3.4, we answer the question in the affirmative in both the binary classification and regression frameworks.  In each framework, we demonstrate sensor decision rules and fusion rules that are universally consistent and connect the results to known work in nonparametrics.

In this model, each sensor's decision rule can be viewed as a selection of one of three states;  abstain, vote and send 1, and vote and send 0.  The option to abstain essentially allows the sensors to convey slightly more information than the one bit that is assumed to be physically transmitted to the fusion center.  With this observation, these results can be interpreted as follows: $\log_2(3)$ bits per sensor per classification is sufficient for universal consistency to hold for both distributed classification and regression \textit{with abstention}.   

In this view, it is natural to ask whether these $\log_2(3)$ bits are necessary.  Can consistency results be proven at lower bit rates?  Consider a revised model, precisely the same as above, except that in response to the fusion center's request for information, each sensor must respond with 1 or 0;  abstention is not an option and thus, each sensor responds with exactly one bit per classification. The same questions arise:  are there rules for which universal consistency results hold in distributed classification and regression \textit{without abstention}?

In Section 3.3 and 3.5, we study distributed classification and regression in communication without abstention.  We observe that universally consistent networks can be designed in the classification regime;   through a negative result, we observe that universal consistency it is not achievable in the regression framework.

\subsection{Distributed Classification with Abstention}
In this section, we show that the universal consistency of distributed classification with abstention follows immediately from Stone's Theorem and the classical analysis of naive kernel classifiers.  Recall, $\mathcal{Y}=\{0, 1\}$ and for each $i\in\{1,...,n\}$, the training datum $(X_i, Y_i)\in D_n$ is received by a distinct member of a network of $n$ sensors.  

To answer the question of whether a universally consistent network can be devised, let us construct one natural choice.  Let \begin{equation}
\delta_{ni}(x)= \left\{%
\begin{array}{ll}
    Y_i, & {\rm if\,\,} X_i\in B_{r_n}(x)\\
     {\rm abstain}, & {\rm otherwise}
\end{array}%
\right.
\end{equation}
and
\begin{equation}
g_n(x)= \left\{%
\begin{array}{ll}
    1, & {\rm if\,\,}\frac{\sum_{i=1}^{n}\delta_{ni}(x) 1_{\{\delta_{ni}(x)\neq {\rm abstain}\}}}{\sum_{i=1}^{n}1_{\{\delta_{ni}(x)\neq {\rm abstain}\}}} \geq \frac{1}{2}\\
     0, & {\rm otherwise}
\end{array}%
\right.\,,
\end{equation}
so that $g_n(x)$ amounts to a majority vote fusion rule.  With this choice, it is straightforward to see that the net decision rule is equivalent to the plug-in kernel classifier rule with the naive kernel.  Indeed, 
\begin{equation}
g_n(x)= \left\{%
\begin{array}{ll}
    1, & {\rm if\,\,} \frac{\sum_{i=1}^{n}Y_i 1_{B_{r_n}(x)}(X_i)}{\sum_{i=1}^{n} 1_{B_{r_n}(x)}(X_i)}\geq \frac{1}{2}\\
     0, & {\rm otherwise}
\end{array}%
\right.\,.
\end{equation}
With this equivalence, the universal consistency of the network follows from Stone's Theorem applied to naive kernel classifiers.  With $L_n = \mathbf{P}\{g_n(X)\neq Y\,| D_n\}$, the probability of error of the network conditioned on the random training data, we state this known result without proof as Theorem 1.

\begin{thm}{}
If, as $n\rightarrow\infty$, $r_n\rightarrow 0$ and $(r_n)^d n\rightarrow\infty$, then $\mathbf{E}\{L_n\}\rightarrow L^{*}$ for all distributions $\mathbf{P}_{XY}$. \end{thm}

\subsection{Distributed Classification without Abstention}
As noted in Section 3.1, given the results of the last section, it is natural to ask whether the communication constraints can be tightened.  Let us consider the second communication model in which the sensors cannot choose to abstain.  In effect, each sensor communicates one bit per decision.  Recall, ${\mathcal{Y}}=\{0,1\}$ and we again consider whether universally Bayes-risk consistent schemes exist for the network.

Let $\{Z_{n,\frac{1}{2}}\}_{n=1}^{\infty}$ be a family $\{0, 1\}$-valued random variables such that $\mathbf{P}\{Z_{n,\frac{1}{2}}=1\} = \frac{1}{2}$.  Consider the randomized sensor decision rule specified as follows:
\begin{equation}
\delta_{ni}(x)= \left\{%
\begin{array}{ll}
    Y_i, & {\rm if\,\,} X_i\in B_{r_n}(x)\\
     Z_{i, \frac{1}{2}}, & {\rm otherwise}
\end{array}%
\right.\,.
\end{equation}
That is, the sensors respond according to their training data if $x$ is sufficiently close to $X_i$.  Else, they simply ``guess", flipping an unbiased coin.  

A natural fusion rule is the majority vote:
\begin{equation}
g_n(x)= \left\{%
\begin{array}{ll}
    1, & {\rm if\,\,} \frac{1}{n}\sum_{i=1}^{n}\delta_{ni}(x) > \frac{1}{2}\\
     0, & {\rm otherwise}
\end{array}%
\right.\,.
\end{equation}
Modifying our convention slightly, let $D_n=\{(X_i, Y_i, Z_{i, \frac{1}{2}})\}_{i=1}^{n}$.   Define 
\begin{equation}\label{riskofrandomrule}
L_n = \mathbf{P}\{g_n(X)\neq Y\, | D_n\}\,.
\end{equation}
That is, $L_n$ is the conditional probability of error of the majority vote fusion rule conditioned on the randomness in sensor training and sensor decision rules.  Assuming a network using the described decision rules, Proposition 1 specifies sufficient conditions for consistency.

\begin{prop}
If, as $n\rightarrow\infty$, $r_n\rightarrow 0$ and $(r_n)^d\sqrt{n}\rightarrow\infty$, then $\mathbf{E}\{L_n\}\rightarrow L^{*}$ for all distributions $\mathbf{P}_{XY}$.
\end{prop}

Yet again, the conditions of the proposition strike a similarity with consistency results for kernel classifiers using the naive kernel.  Indeed, $r_n\rightarrow 0$ ensures the bias of the classifier decays to zero.  However, $\{r_n\}_{n=1}^{\infty}$ must not decay too rapidly.  As the number of sensors in the network grows large, many, indeed most, of the sensors will be ``guessing" for any given prediction; in general, only a decaying fraction of the sensors will respond with useful information.  In order to ensure that these informative bits can be heard through the noise introduced by the guessing sensors, $(r_n)^d \sqrt{n}\rightarrow\infty$.  Note the difference between the result for naive kernel classifiers where $(r_n)^d n\rightarrow\infty$ dictates a sufficient rate of convergence for $\{r_n\}_{n=1}^{\infty}$.

To prove this result, we show directly that the expected probability of misclassification converges to the Bayes rate.  This is unlike techniques commonly used to demonstrate the consistency of kernel classifiers, etc., which are so-called ``plug-in" classification rules.  In those settings, it suffices to show that the rules are based on consistent estimates of the \textit{a posteriori} probabilities $\mathbf{P}\{Y=i\,|X\}$, $i\in\{0,1\}$. However, for this model, we cannot estimate the \textit{a posteriori} probabilities directly; the proof resorts to margin-based analysis \cite{PreKulPoo04b}.  These comments foreshadow the negative result of Section 3.5.

\subsection{Distributed Regression with Abstention}
Let us now move to a regression setting in which we focus on estimating a real-valued function in a bandwidth starved environment.  The model remains the same except that $\mathcal{Y}=\bbbr$; that is, $Y$ is a $\bbbr$-valued random variable and likewise, sensors receive real-valued training data labels, $Y_i$.   As in Section 3.2,  for each prediction, sensors are allowed to transmit one bit of information and they have the ability to abstain.  To demonstrate that consistency can be achieved, let us devise candidate rules.  

For each integer $n$, let $\{Z_{n,\theta}\}_{\theta\in[0,1]}$ be a family of random $\{0,1\}$-valued random variables parameterized by $[0,1]$ such that for each $\theta\in[0,1]$, $Z_{n, \theta}$ is Bernoulli with parameter $\theta$.

Let $\{c_n\}_{n=1}^{\infty}$ and $\{r_n\}_{n=1}^{\infty}$ be arbitrary sequences of real numbers such that $c_n\rightarrow\infty$ and $r_n\rightarrow 0$ as $n\rightarrow\infty$.  Let local sensor decision algorithm $\delta_{ni}(x)$ be defined as:
\begin{equation}\label{agentdr}
\delta_{ni}(x)= \left\{%
\begin{array}{ll}
    Z_{i, \frac{1}{2c_n}Y_i+\frac{1}{2}}, & {\rm if\,\,} x\in B_{r_n}(X_i)\,\, {\rm and}\,\, |Y_i|\leq c_n\\
    Z_{i,\frac{1}{2}}, & {\rm if\,\,} x\in B_{r_n}(X_i) \,\,{\rm and} \,\,|Y_i|> c_n\\
    {\rm abstain}, & {\rm otherwise}  \\
\end{array}%
\right.\,,
\end{equation}
for $i=1,...,n$.  In words, the sensors choose to vote if $X_i$ is close enough to $X$; to vote, they flip a biased coin, with the bias determined by  $Y_i$ and the size of the network, $n$.

Let us define the fusion rule:
\begin{eqnarray}\label{fusion}
\hat{\eta}_n(x) & = & 2 c_n\Big{(}\frac{\sum_{i=1}^n \delta_{ni}(x)1_{\{\delta_{ni}(x)\neq {\rm abstain}\}}}{\sum_{i=1}^n 1_{\{\delta_{ni}(x)\neq {\rm abstain}\}}} - \frac{1}{2}\Big{)}\,.
\end{eqnarray}
In words, the fusion rule shifts and scales the average vote.

Define $L_n = \mathbf{E}\{|\hat{\eta}_n(X)-Y|^2\,|D_n\}$ with the expectation taken over $X$, $D_n=\{(X_i, Y_i)\}_{i=1}^{n}$, and the randomness introduced in the sensor decision rules.  Assuming a network using the described decision rules, Proposition 2 specifies sufficient conditions for consistency.
 \begin{prop}
 Suppose $\mathbf{P}_{XY}$ is such that $\mathbf{P}_X$ is compactly supported and $\mathbf{E}\{Y^2\}<\infty$. If, as $n\rightarrow\infty$,
 \begin{enumerate}
\item $c_n\rightarrow\infty$\,,
\item $r_n\rightarrow 0$, and
\item $\frac{c_n^2}{n (r_n)^d}\rightarrow 0$, 
\end{enumerate}
then $\mathbf{E}\{L_n\}\rightarrow L^{\star}$. 
 \end{prop} 
 Those familiar with the classical statistical pattern recognition literature will find the style of proof very familiar; special care must be taken to demonstrate that the variance of the estimate does not decrease too slowly compared to  $\{c_n\}_{n=1}^{\infty}$ and to show that the bias introduced by the ``clipped" sensor decision rules converges to zero.   Note that the divergent scaling sequence $\{c_n\}_{n=1}^{\infty}$ is required for the general case when there is no reason to assume that $Y$ has a known bound; in general, any decision rule which obeys the communication constraints will require a scaling sequence ``like" $\{c_n\}_{n=1}^{\infty}$ .  If, instead, $|Y|\leq B$ a.s. for some known $B>0$, it suffices to let $c_n=B$ for all $n$.   More generally, the constraint regarding the compactness of $\mathbf{P}_{X}$ can be weakened.  For a more detailed discussion, we refer the reader to \cite{PreKulPoo04b}.

\subsection{Distributed Regression without Abstention}
Finally, let us consider the communication model from Section 3.3 in the regression setting.  Now, $\mathcal{Y}=\bbbr$; sensors will receive real-valued training data labels $Y_i$.  When asked to respond with information, they will reply with either $0$ or $1$.   We will argue that universal consistency is not achievable in this one bit regime.

Let $A=\{a:\bbbr^d\times\bbbr^d\times\bbbr\rightarrow [0,1]\}$.  That is, $A$ is the collection of functions mapping $\bbbr^d\times\bbbr^d\times\bbbr$ to $[0,1]$.  For every sequence of functions $\{a_n\}_{n=1}^{\infty}\subseteq A$, there is a corresponding sequence of randomized sensor decision rules $\{\delta_{ni}(X)\}_{n=1}^{\infty}$, specified by
\begin{equation}\label{agentdr2}
\delta_{ni}(x)=Z_{i, a_n(x,X_i, Y_i)}\,,
\end{equation}
for $i\in\{1,...,n\}$.   Let us consider the set of sensor decision rules so specified.  Note that as before, these sensor decision rules are allowed to depend on $n$ and satisfy the same constraints imposed on the decision rules in the classification framework of Section 3.3.  

A  fusion rule consists of a sequence of functions $\{\hat{\eta}_n\}_{n=1}^{\infty}$ mapping $\bbbr^d\times\{0, 1\}^n$ to $\mathcal{Y}=\bbbr$.  To proceed, we require some regularity on $\{\hat{\eta}_n\}_{n=1}^{\infty}$.  
We impose two natural constraints:  (i) the fusion rule must be permutation invariant to the sequence of bits sent from the $n$ sensors and (ii) the fusion rule should be Lipshitz in the average Hamming distance, i.e., there exists some constant $C$ such that
\begin{equation}\label{continuity}
|\hat{\eta}_n(x, b_1) - \hat{\eta}_n(x, b_2)| \leq C \frac{1}{n}{\sum_{i=1}^n |b_{1i} - b_{2i}|}
\end{equation}
for all bit strings $b_1, b_2\in\{0,1\}^n$, all $x\in\bbbr^d$, and every $n$. 

As usual, we will consider $L_n = \mathbf{E}\{|\hat{\eta}_n(X)-Y|^2\,|D_n\}$ as the performance metric;  here, the expectation is taken over $X$ and any randomness introduced in the sensor decision rules themselves. The main result is as follows.

\begin{prop}
For every sequence of sensor decision rules $\{\delta_{n}(x)\}_{n=1}^{\infty}$ specified according to (\ref{agentdr2}) with a pointwise converging sequence of functions $\{a_n\}_{n=1}^{\infty}\subseteq A$, there is no permutation invariant fusion rule $\{\hat{\eta}_n\}_{n=1}^{\infty}$ satisfying (\ref{continuity}) such that
\begin{equation}
\lim_{n\rightarrow\infty}\mathbf{E}\{L_n\}=L^{\star}
\end{equation}
universally.
\end{prop}

The proof in \cite{PreKulPoo04b} proceeds by using $\{a_n\}_{n=1}^{\infty}$ to specify two random variables $(X, Y)$ and $(X^{\prime}, Y^{\prime})$ with $\eta(x)=\mathbf{E}\{Y\,| X=x\}\neq\mathbf{E}\{Y^{\prime}\,| X^{\prime}=x\}=\eta^{\prime}(x)$.  Asymptotically, however, the fusion center's estimate will be indifferent to whether the sensors are trained with random data distributed according to $\mathbf{P}_{XY}$ or $\mathbf{P}_{X^{\prime}Y^{\prime}}$.  This observation will contradict universal consistency and complete the proof.

\section{Learning with Specialists}

In the previous section, the learning problem was divided amongst sensors by first sampling i.i.d. data according to the underlying unknown probability distribution;  the data was then distributed amongst the sensors.  In this section, we present a second model where sensors are first assigned random, local subsets of the observation space.  Then, the sensors become specialized in these regions by observing i.i.d. training examples according the underly distribution, but which are constrained to fall within the sensor's local region of specialization.  As before, the sensors are constrained in the way they can communicate.  With this new sampling process, we could naturally consider the four cases corresponding to classification and regression in communication models with and without abstention.  However, to illustrate this model and to understand the underlying fundamentals, we will consider only the
case of classification with abstention.  Other cases can be considered similarly. Thus, when the network observes a new  observation, the sensors can respond with up to one bit of information, retaining the flexibility to abstain.  The fusion center combines this information to make a prediction.

Though the distinction between the models may appear subtle, the difference is essential and is motivated by a different set of sensor network applications and notions of being distributed.  The model is perhaps most relevant in applications such a field estimation or other scenarios where a dimension of the observation space describes the position of a sensor;  here expertise, rather than data, is distributed.  Random assignment of regions of specialization can model random dispersal of sensors about an environment. Our model will be posed more generally as this allows us to understand the fundamental differences between this model and its classical counterpart.
\subsection{The Model \& Main Result}
Consider a special case of a general formulation in \cite{PreKulPoo04c}.  Let $\mathcal{X}=[0,1]^d$ and $\mathcal{Y}=\{0,1\}$.  Suppose $n$ sensors are randomly assigned subsets of $\mathcal{X}$ in which to specialize; i.e., let $\{\Theta_i\}_{i=1}^n$ be a collection of i.i.d. random variables \emph{uniformly distributed} across $\mathcal{X}$ so that $B_{r_n}(\Theta_i) = \{x\in\mathcal{X}:\,||x-\Theta_i||_2\leq r_n\}$ is the local \emph{region of specialization} for sensor $i$.  Each sensor samples a training datum according to the distribution $\mathbf{P}\{X,Y\,|X\in B_{r_n}(\Theta_i)\}$.    That is, sensor $i$ receives one labeled training example $(X_i, Y_i)$ distributed according to $\mathbf{P}_{XY}$, conditioned on $X_i$ being in the sensor's region of specialization, $B_{r_n}(\Theta_i)$.

After training, the fusion center observes $X\sim{\mathbf{P}}_X$ and broadcasts it to the network in a request for information; sensors respond with one bit according to the following rule:

\begin{equation}\label{dr}
\delta_i(X)= \left\{%
\begin{array}{ll}
    Y_i, & X\in B_{r_n}({\Theta_i}) \\
    {\rm abstain}, & {\rm otherwise}  \\
\end{array}%
\right..
\end{equation}
That is, the sensors respond with one bit according to their training data label as long as the new observation $X$ falls within its region of specialization.  Otherwise, they do not respond.

A fusion center combines this information with a majority vote:
\begin{equation}\label{fusion}
g_{n}(X)=\left\{%
\begin{array}{ll}
    1 , & \frac{1}{\Lambda(X)}\sum_{i=1}^{n}\delta_{i}(X)\geq\frac{1}{2} \\
    0 , & {\rm otherwise} \\
\end{array}%
\right.,
\end{equation}
with $\Lambda(x) = \sum_{i=1}^n\mathbf{1}_{\{x\in B_{r_n}(\Theta_i)\}}$.  

For example, if $d=2$, we might regard $\mathcal{X}$ as a cityscape so that each $x\in\mathcal{X}$ is a location and $y\in\mathcal{Y}$ is a binary label describing whether a toxin is present.  With this example, $X$ is a random location of interest to an analyst and $Y$ is a random realization of the toxicity; the sensors randomly deployed across $\mathcal{X}$ form a sensor network that learns toxicity as a function of position.

The overriding question is whether this network can be designed to be universally consistent.  Let $L_n = \mathbf{P}\{g_n(X)\neq Y\,|\{\Theta_{ni}, X_i, Y_i\}_{i=1}^n \}$ denote the expected probability of error of $g_n$ conditioned on random sensor specialization and training.  Can the network choose $r_n$ such that ${\mathbf{E}}\{L_n\}\rightarrow L^{\star}$ universally?  This question is answered in the following proposition.

\begin{prop}  If $r_n\rightarrow 0$ and $(r_n)^d n\rightarrow\infty$, then ${\mathbf{E}}\{L_n\}\rightarrow L^{*}$ for all distributions $\mathbf{P}_{XY}$. \end{prop}

Though the proposition resembles classic results for the universal consistency of kernel classifiers \cite{DevGyoLug96}, the key difference is the process from which training data is sampled.   In traditional models, the sampling process by which training data is received is independent and identically distributed according to underlying unknown probability distribution.  In the current model, though i.i.d, training examples are generated from a distribution which first depends on a random allocation of sensors in the observation space.  This distribution is in general different than the underlying distribution ${\mathbf{P}}_{XY}$ and moreover, it evolves with $n$ as the sensors grow dense in the observation space.  Though this difference may appear to be technical, it is fundamental, arises from the study of distributed learning, and precludes applying previously known results in nonparametrics.  In \cite{PreKulPoo04c}, a more general formulation results in a theorem that quantifies the difference through a universal relationship between the random specialization of the sensors and probability distributions defined on compact observation spaces.  The proof is of a similar style to classical consistency results, taking into account such differences.

\section{Conclusions}
In this paper, we have described several models for distributed
learning within communication constraints, and we have discussed issues
of consistency within them.   These models are of particular interest
for potential applications in wireless sensor networks, given the intent
of and constraints on these networks.  It is anticipated that further research
will lead to such applications.

\small
\bibliography{Allerton-bib}
\bibliographystyle{IEEEtranS}

\end{document}